\documentclass[prb,aps,amsmath,preprint,]{revtex4-1}
\usepackage{graphicx}

\usepackage{dcolumn}

\usepackage{bm}
\usepackage{subfigure}
\usepackage{ltxtable, tabularx, longtable}
\usepackage{color}

\begin{document}

\title{Structural, vibrational, and quasiparticle band structure of 1, 1 - diamino-2, 2 - dinitroethelene from ab-initio calculations.}

\author{S. Appalakondaiah and G. Vaitheeswaran$^*$}
\affiliation{Advanced Centre of Research in High Energy Materials (ACRHEM),\\
University of Hyderabad, Prof. C. R. Rao Road, Gachibowli, Andhra Pradesh, Hyderabad- 500 046, India.}

\author{S. Leb\`egue}
\affiliation{Laboratoire de Cristallographie, R\'esonance Magn\'etique et Mod\'elisations (CRM2, UMR CNRS 7036), Institut Jean Barriol, Universit\'e de Lorraine, BP 239,\\
 Boulevard des Aiguillettes, 54506 Vandoeuvre-l\`es-Nancy, France.}

\date{\today}
\begin{abstract}
The effect of pressure on the structural and vibrational properties of the layered molecular crystal 1,1-diamino-2,2-dinitroethelene (FOX-7) are explored
by first principles calculations. We observe significant changes in the calculated structural properties with different corrections
for treating van der Waals interactions to Density Functional Theory (DFT), as compared with standard DFT functionals.
In particular, the calculated ground state lattice parameters, volume and bulk modulus obtained with Grimme's scheme are found to agree well with experiments.
The calculated vibrational frequencies demonstrates the dependence of the intra and inter-molecular interactions in FOX-7 under pressure.
In addition, we also found a significant increment in the N-H...O
hydrogen bond strength  under compression. This is explained by the change in bond lengths between nitrogen, hydrogen and oxygen atoms, as well as calculated IR spectra under pressure.
Finally, the computed band gap is about 2.3 eV with GGA, and
is enhanced to 5.1 eV with the  GW approximation, which reveals the importance of performing quasiparticle calculations in high energy density materials.

\end{abstract}
\maketitle
\section{Introduction}

1, 1 - diamino-2, 2 - dinitroethelene (C$_2$H$_4$N$_4$O$_4$, commonly known as FOX-7) is a layered molecular crystal, which belongs to the high energy materials (HEMs) family and has generated considerable interest due to its low sensitivity and high thermal conductivity.
It is an attractive HEM among the class of CHNO based materials due to its extreme energetic characteristics such as high performance and sensitivity: FOX-7 has a comparable performance
 to related HEMs like 1,3,5-Trinitroperhydro-1,3,5-triazine (RDX) and  Octahydro-1,3,5,7-tetranitro-1,3,5,7-tetrazocine (HMX) and it is more sensitive than
 1,3,5-triamino-2,4,6-trinitrobenzene (TATB)\cite{Mathieu,Anniyappan}.
 Starting from the synthesis of this
compound in 1998\cite{Nikolai}, several studies were carried out to investigate its structural properties from ambient\cite{Ulf} to extreme conditions\cite{Peris,Kempa,Evers,Pravica}, its
 thermal behavior with decomposition mechanism\cite{Xue, Gao} and vibrational spectroscopy (IR and Raman) at different pressures and temperatures\cite{Peris,pravica,Bishop}. Beeman and
 Ostrak\cite{Ulf} determined the crystal structure of FOX-7 to be monoclinic with the space group P2$_1$/n, which contains four molecules (56 atoms) per unit cell. Later Peris
 et al\cite{Peris} studied its structural properties and Raman spectra under the influence of non-hydrostatic pressure and found a quasi-amorphous phase above 4.5 GPa. However,
 they found that no lattice transformations could be observed up to 8 GPa  under hydrostatic conditions.  Recent experiments at  different temperatures and pressures found three
 possible phase transitions (at 2 GPa, 5 GPa and above 10 GPa) in mid and far IR regions \cite{pravica,Bishop}. Pravica et al\cite{pravica} reported the strengthening of the
 hydrogen bond and softening of NH$_2$ stretching frequencies at high pressure by using IR experiments. Very recently, Dreger et al\cite{gupta} studied Raman spectra under isothermal compression (up to 15 GPa) and isobaric heating (up to 500K), and claimed that two phase transitions at 2 and 4.5 GPa.
  Apart from this, FOX-7 has shown three (possibly four) solid polymorphic phases as a function of temperature\cite{Kempa,Evers,margaret}. The $\alpha$ phase ( monoclinic,P2$_1$/n, Z=4) appears to be most stable under ambient conditions, whereas it transforms to $\beta$ phase (orthorhombic, P2$_1$2$_1$2$_1$, Z=4) beyond 378 K at ambient pressure and further transorms to $\gamma$ phase (monoclinic,P2$_1$/n, Z=8) upon  heating above 448 K. Above this, by lowering the temperature with increasing pressure, FOX-7 incompletely converts to $\alpha$ phase\cite{margaret}.
 On the theoretical side, Kukilja and co workers reported shear-strain
 effects on decomposition mechanism,  mechanical compression, electronic excitations with modifications in molecular and crystalline forms of FOX-7 with different methods
\cite{Kuklja1,Kuklja2,Kuklja3,Kuklja4,Kuklja5,Kuklja6,Kuklja7,Kuklja8,Kuklja9,Kuklja10,Kuklja11,Kuklja12,Kuklja13}. A few density functional theory (DFT) calculations were
 performed earlier to describe the structural properties at ambient conditions, the phase stability up to 4 GPa under hydrostatic pressure, and the electronic properties and decomposition
mechanism\cite{Asta,Sorescu,Hai,Hu,Zhao,Zneng}. Recently, van der Waals (vdW) corrected DFT studies at ambient conditions were reported\cite{dftd,Landeville}, but until now
 there is no theoretical work to understand the vibrational and excited properties under hydrostatic pressures.

 \begin{figure*}
\centering
\subfigure[]{\includegraphics[width=2.0in,height=2.0in]{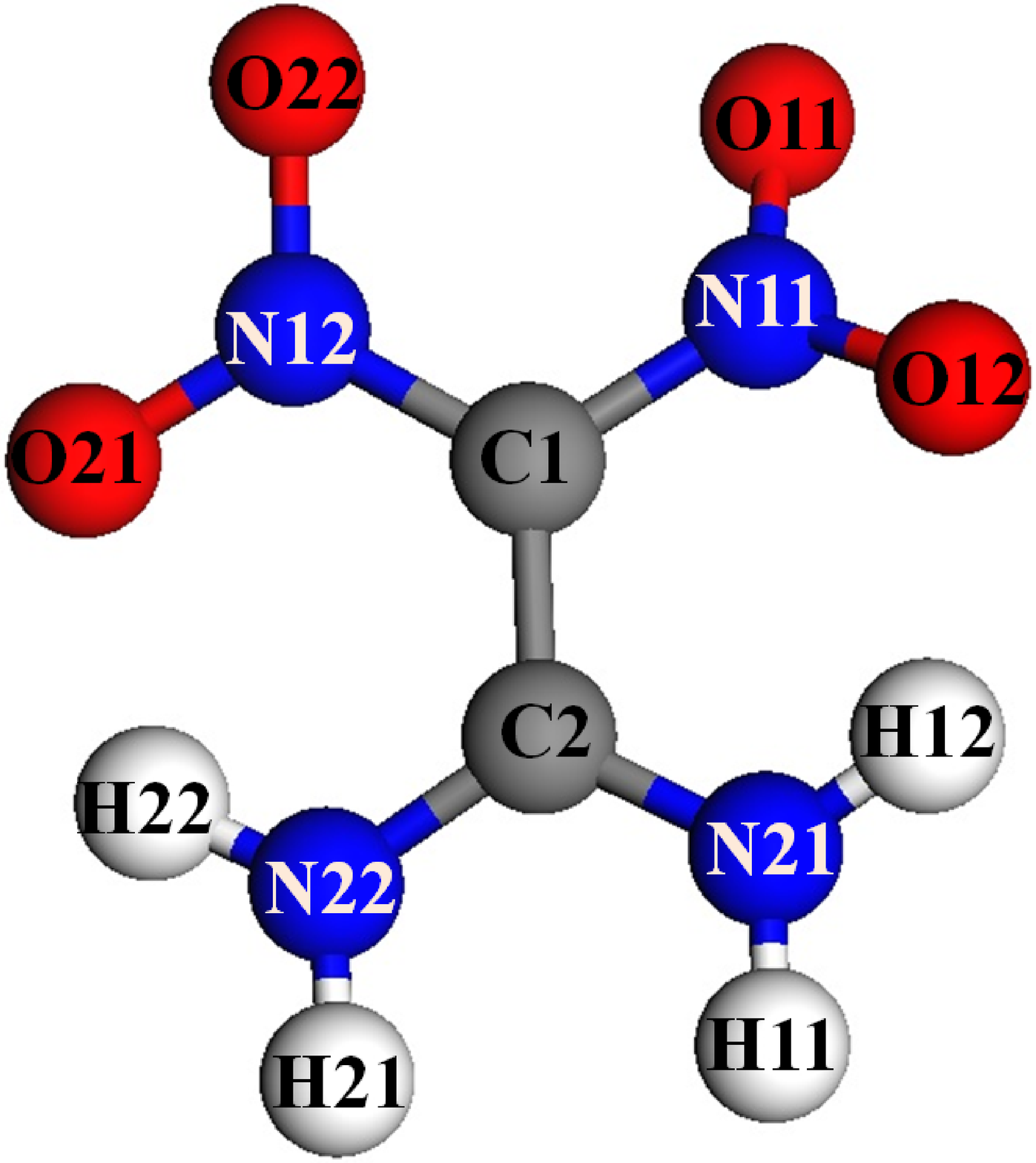}}
\subfigure[]{\includegraphics[width=4.3in,height=3.5in]{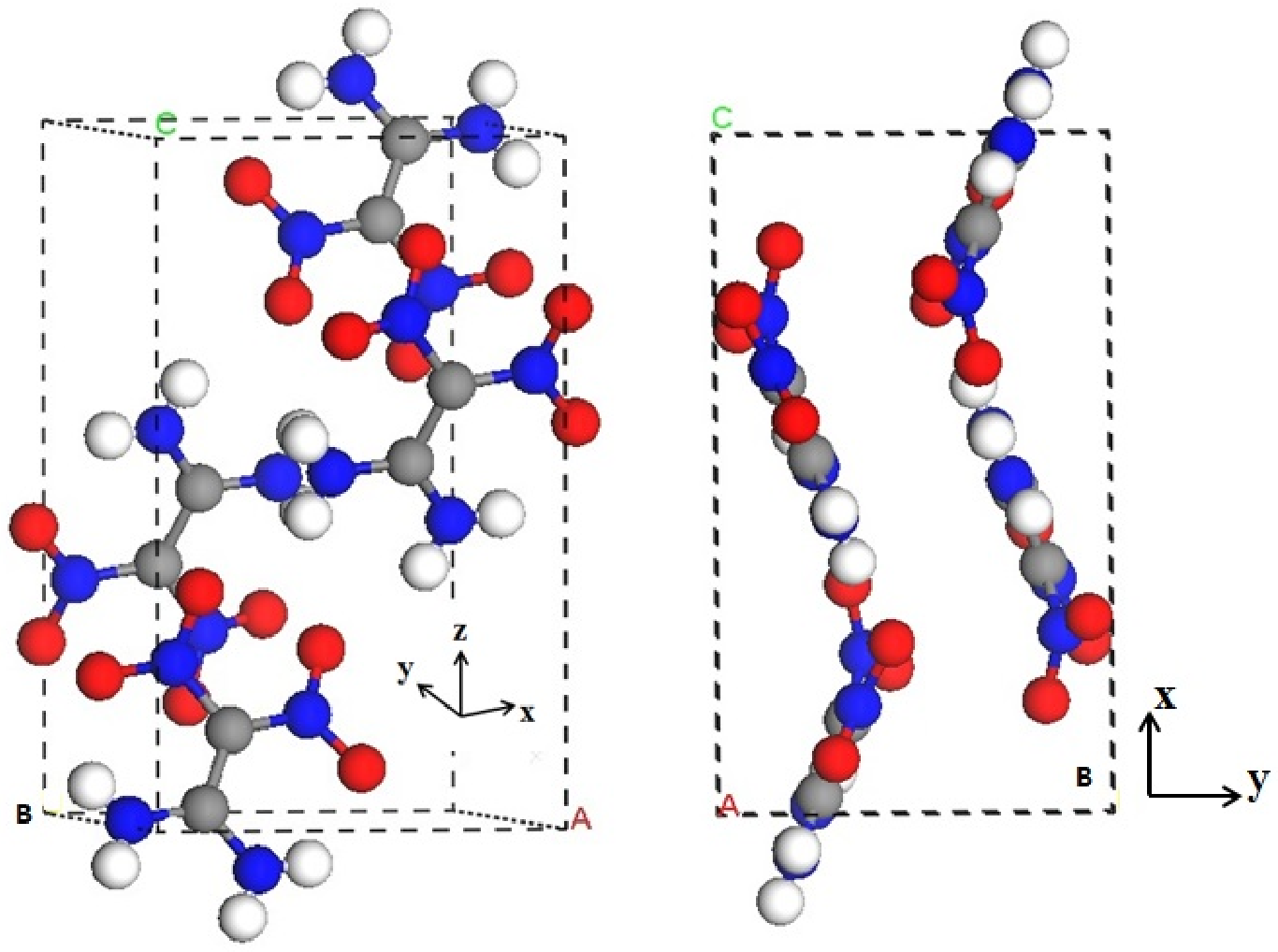}}
\caption{(Color online) Experimental structure of FOX-7: (a) single molecule (b) complete crystal structure along different directions.
}
\end{figure*}

\par Since FOX-7 is a layered molecular system, vdW forces between the layers are playing a crucial role in holding the structure, together with
 strong hydrogen bonding within the layers due to the close contact of the NH$_2$ group of one molecule with the NO$_2$ group of the adjacent molecule.
 vdW forces are not taken into account in standard DFT approximations, such as the Local Density Approximation (LDA) or
 the Generalized Gradient Approximation (GGA) due to the fact that they are non-local in nature.
 In the present manuscript, we show that vdW forces cannot be ignored in order to study the structural and vibrational properties of crystalline FOX-7,
  and we took them into account by using and comparing various DFT+D methods.
  Also, we have studied the electronic bandstructure of FOX-7 with the GW approximation to know the band gap value accurately.
 The rest of the paper is organized as follows: in section II, we describe the computational techniques that we used, while the results and discussions
 are presented in section III. Finally, a brief conclusion is given in section IV.

\section{Computational details}
Our DFT calculations were performed with the plane wave pseudopotential method as implemented in the CAmbridge Serial Total Energy Package (CASTEP) code\cite{Payne, Segall}.
The calculations were done using Vanderbilt ultrasoft pseudopotentials for all atoms\cite{Vanderbilt}, and
 the LDA in the scheme of CAPZ (Ceperley and Alder as parametrized by Perdew and Zunger)\cite{PPerdew}), GGA  with PBE (Perdew-Burke-Ernzerhof)\cite{Perdew} and
 PW91 (Perdew and Wang)\cite{PW91} functionals were used as the exchange and correlation functionals. To correct DFT for the missing vdW interaction, we have used three types of
  corrections such as the Grimme (G06)\cite{Grimme} and the Tkatchenko and Scheffler (TS)\cite{TS}  corrections to PBE, and the Ortmann, Bechstedt, and Schmidt (OBS)\cite{OBS}
  correction to the PW91 functional.
A plane wave kinetic energy cutoff of 540 eV was used, and the first Brillouin zone  was sampled on a regular Monkhorst-Pack\cite{Monkhorst} grid  with a minimum
 spacing of 0.04\AA$^{-1}$. The geometry optimization of the system has been achieved by relaxing the forces ($<$ 1e-5 eV/\AA), the total energy ($<$ 5e-5 eV/atom),
 and the stress tensor ($<$ 0.02 GPa). The vibrational frequencies have been calculated from the response to small atomic displacements within the linear
 response approach as implemented in CASTEP code.  An accurate band gap for FOX-7 was obtained with the GW approximation\cite{Hedin1,Hedin2,seb1,seb2} as implemented in the Vienna Ab initio Simulation Package\cite{Kresse}.
 To obtain convergence, we used 250 bands for the summation over the bands in the polarizability and the self-energy formulas, and
 the polarizability matrices were calculated up to a cut-off of 150 eV.

\section{Results and discussions}
\subsection{Structural properties}

We have started our study by calculating the ground state geometry of FOX-7 crystal using the various functionals mentioned above.
 As presented in Table I, we found a large difference in the calculated equilibrium volumes with either the LDA (underestimated by around 10\%) or with the GGA (overestimated
 by around 28\% with PW91 and by around 27\% with PBE). Also, by comparing the three calculated lattice parameters of the monoclinic FOX-7 unit cell, we observe
 a larger deviation for the {\it b} lattice parameter as compared with other the two parameters {\it a} and {\it c}.
 From the experimental crystal structure of FOX-7 (see Fig 1), it can be seen that the {\it b} crystallographic direction corresponds mainly to the direction
  of stacking of the layers which are
 binded by vdW interactions\cite{Peris}. This inconsistency between the experimental data and the calculated results obtained with standard DFT functionals
 can be corrected when using the DFT+D methods: the calculated volume using the GGA functional corrected  with the OBS method is underestimated by 0.6\%, the
 TS method overestimates the same by 2.7\% while the G06 correction improves the volume to a greater extent at almost ~0.1\%  less than the experimental value.
 Therefore for further calculations, we adopted the G06 correction to PBE (here after labeled as GGA+G06) to study properties under high pressure.
 Notice that our calculated structural
 parameters are in good agreement with previous theoretical studies on this material\cite{Zhao,dftd,Landeville}.

\begin{table}[]
\caption{The calculated ground state properties of monoclinic FOX-7 at ambient pressure. a, b and c are the lattice parameters (in \AA), monoclinic angle ($\beta$),  V (\AA{$^3$}) the volume of the unit cell.
}
\begin{ruledtabular}

\begin{tabular}{cccccccccccccccccccc}
axis   & LDA   & PW91  & PBE   & OBS & TS & G06 & Exp\cite{Ulf}    \\ \hline
 a      & 6.75  & 7.24  & 7.23  & 6.93     & 6.99   & 6.99    & 6.941  \\
 b      & 6.21  & 7.88  & 7.78  & 6.56     & 6.66   & 6.52    & 6.569  \\
 c      & 11.05 & 11.63 & 11.65 & 11.29    & 11.38  & 11.31   & 11.315 \\
 $\beta$  & 90.62 & 92.14 & 92.02 & 90.98    & 91.41  & 91.23   & 90.55  \\
 V      &462.51 &662.45 & 654.22& 512.65   & 529.64 & 515.45  & 515.89 \\


 \end{tabular}
\end{ruledtabular}
\end{table}

\par Then, the equation of state (EoS) of FOX-7 was calculated in a pressure range from 0 to 10 GPa with a step size of 1 GPa.
The obtained lattice parameters and the corresponding volumes are shown in Fig. 2.
At low pressures, large deviations are observed with LDA and PBE, whereas under pressure all the three functionals shows a similar behavior,
 which highlight the fact that van der Waals interaction becomes less important under large pressure.
The deviation of our calculated lattice parameters at 0 GPa and 4 GPa with available experiments (up to 3.9 GPa) are listed in Table II.
By comparing the three lattice parameters, the reduction of the {\it b} lattice
parameter under pressure is larger than the other two parameters as clearly observed from the first order pressure coefficients ( $\gamma = \frac{1}{X}\frac{dX}{dP}$, X={\it a}, {\it b}, {\it c}).

\begin{figure*}
\centering
{\includegraphics[width=6in,clip]{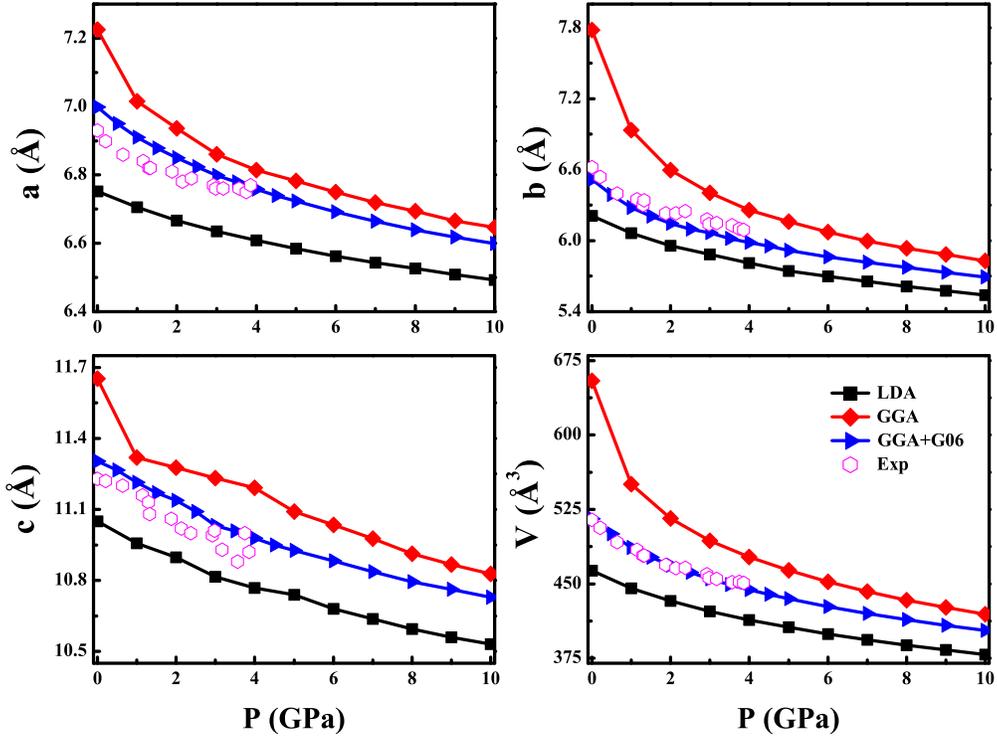}}
\caption{(Color online) Hydrostatic pressure dependence of lattice parameters (a, b, and c) and  crystal volume of FOX-7 up to 10 GPa as calculated within LDA, GGA-PBE and GGA+G06 and as compared with experiments.}
\end{figure*}

The calculated pressure coefficients obtained with either LDA, PBE and GGA+G06 are presented in table III, following an order of {\it b}$>$ {\it c}$>$ {\it a} with all three methods,
which can be understood as FOX-7 having the largest compressibility along the {\it b}-axis, as reported also experimentally\cite{Peris}. Then, the EoS of FOX-7 is determined by fitting the P-V
data to a second order Birch-Murnaghan (B-M) EoS. We found that the bulk moduli (see Table III) obtained with the LDA or the GGA is in poor agreement with experiments, while the one obtained
with GGA+G06 is in very good agreement with the experimental value of 17.6 GPa\cite{Peris}.

\begin{table}[]
\caption{The deviation of the lattice parameters (in \AA) of monoclinic FOX-7 at 0 GPa and 4 GPa with experiments. Here '-' sign indicates an underestimation and '+' indicates an overestimation in comparison with experimental values }
\begin{ruledtabular}
\begin{tabular}{cccccccccccccccccccccccccccc}
XC&&at 0 GPa &&&& at 4 GPa \\ \hline
 &    a   &   b   &   c   &&     a   &   b   &   c \\ \hline
LDA &   -0.18 & -0.41 & -0.19  &&  -0.17 & -0.28 & -0.15 \\
PBE &   +0.30 & +1.16 & +0.42  &&  +0.05 & +0.17 & +0.27 \\
GGA+G06 & +0.07 & -0.10 & +0.07 && +0.01 & -0.11 & +0.06 \\
 \end{tabular}
\end{ruledtabular}
\end{table}

\begin{table}[]
\caption{The calculated first order pressure coefficients (in 10$^{-3}$ \AA$^{-1}$ GPa$^{-1}$)of the lattice parameters and second order bulk moduli (in GPa) of monoclinic FOX-7.}
\begin{ruledtabular}
\begin{tabular}{cccccccccccccccccccc}
XC &    a   &   b   &   c   &B\\ \hline
LDA &   -3.9 & -9.3 & -7.6  &31.26\\
PBE &   -7.9 & -19.8 & -12.2 &8.13\\
GGA+G06 & -3.4 & -6.3 & -5.0&18.43 \\
 \end{tabular}
\end{ruledtabular}
\end{table}

\subsection{Vibrational properties}

We have also studied the vibrational properties of FOX-7 by computing the phonon frequencies at the $\Gamma$ point with the GGA+G06 functional.
The unit cell  contains 56 atoms and has therefore a total of 168 (3 acoustic+165 optical)  modes.
  The optical modes have the following irreducible representation:
  42A$_g$+42B$_g$+41A$_u$+40B$_u$, where A$_g$, B$_g$ modes have inversion symmetry and are Raman active;  A$_u$, B$_u$ are  IR active due to change of sign under inversion symmetry.
  The main characteristics are the
  following:
  the lower frequencies from 22 to 470  cm$^{-1}$ are external modes, i.e. vibration from all atoms in unit cell,
  (a) The internal modes from 530-713  cm$^{-1}$ corresponds to  wagging and rocking motion of the NH$_2$ group
  (b) Mixed motions from the NH$_2$, NO$_2$, C-N groups stretching and N-C-N rotations are observed between 713-832 cm$^{-1}$,
  (c) Individual NH$_2$ group rocking is found from 1006-1053 cm$^{-1}$,
  (d) The modes between 1104-1605 cm$^{-1}$ corresponds to NO$_2$ stretching, NO$_2$ bending and C-C stretching
  (e) the very high frequencies (more than 3200 cm$^{-1}$) are from NH$_2$ symmetric and asymmetric stretching modes.
  Also we have calculated the phonon frequencies of FOX-7 under pressure from 0 to 10 GPa
  and the calculated IR spectra in from 0 to 10 GPa, as shown in Fig 3.
  According to our calculated IR spectra, we found that all
  internal modes from 530-1006 cm$^{-1}$ (Fig. 3(a)) and 1104-1605 cm$^{-1}$ (Fig. 3(b), 3(c)) are shifted towards higher frequencies with increasing pressure.
  On the contrary, the modes above 3200 cm$^{-1}$ (Fig. 3(d)), are shifted to lower values with increasing pressure.
  This behavior illustrates the enhancement of intermolecular bond strength under hydrostatic pressure conditions.
  The obtained lattice modes with respect to external pressure are shown in Fig. 4.

\begin{figure*}
\centering
{\includegraphics[width=6in,clip]{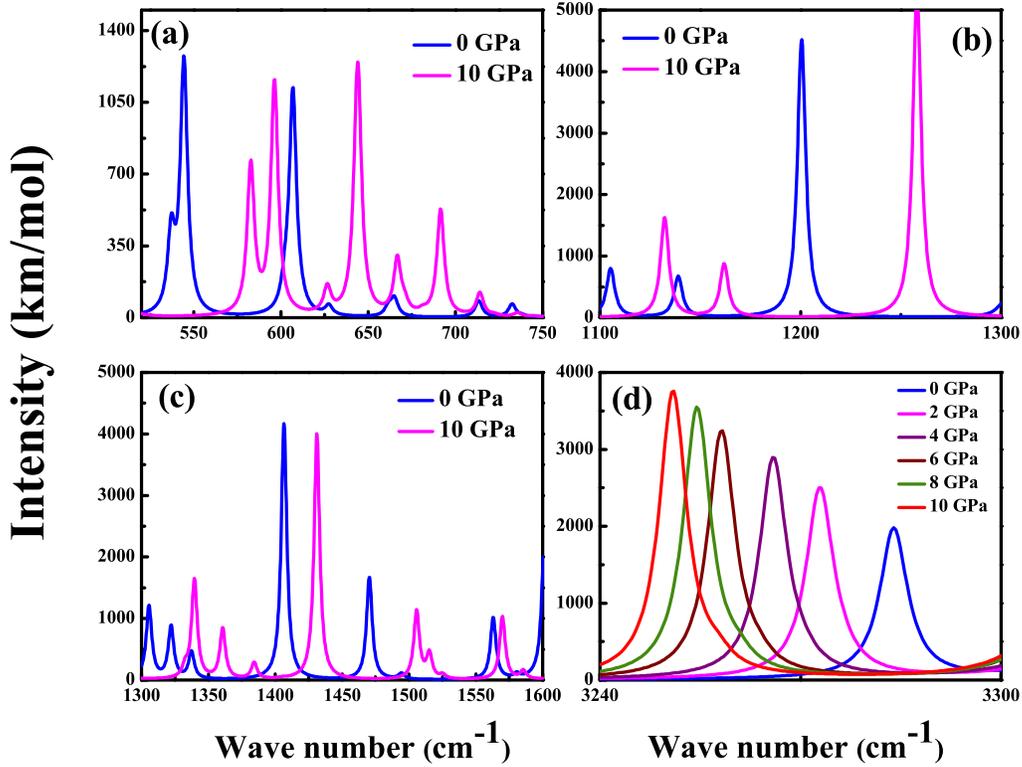}}
\caption{(Color online) IR spectra of FOX-7 under hydrostatic pressure range from 0 to 10 GPa.
}
\end{figure*}

\begin{figure*}
\centering
{\includegraphics[width=6in,clip]{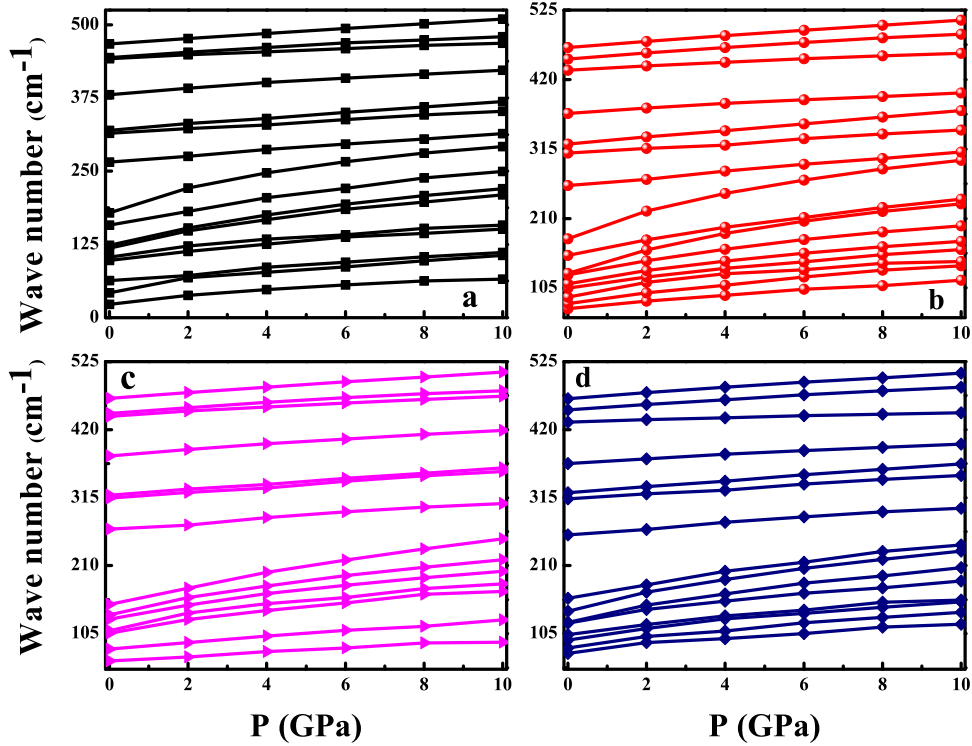}}
\caption{(Color online) Lattice mode frequencies of FOX-7 under hydrostatic pressure range from 0 to 10 GPa. Here a, b, c and d are A$_g$, B$_g$, A$_u$  and B$_u$ modes respectively.
}
\end{figure*}


\subsection{Hydrogen bonding}

From recent experiments on FOX-7\cite{pravica,Bishop}, there is evidence for an increase of the hydrogen bond strength under pressure.
 In order to confirm the nature of  hydrogen bonding under hydrostatic pressure, we investigated the intra- as well as inter- molecular N-H...O
 hydrogen bonds and the IR spectra of FOX-7 under pressure. The intra and inter molecular Donor (Nij)- Hydrogen (Hij), Acceptor (Oij)- Hydrogen (Hij) and
 Acceptor (Oij)- Donor (Nij) bond lengths (here ij notations are defined in Fig. 1(a)) under pressure are shown in Fig 5. All the intramolecular N-H bond lengths (shown in Fig. 5(a))
 decreases except N21-H11. The bond length variations for O-H and N-H are shown in Fig 5(b) and Fig. 5(c) respectively.
Also the O-H and N-O bonds between adjacent molecules are found to  decrease drastically with pressure and some of the O-H
 bond lengths are shorter than the sum of their vdW radii. This indicates a large variation in the intermolecular interactions and affects the strength of hydrogen
 bonding in FOX-7.
 Also, the signature of hydrogen bond was observed from IR spectra in two cases previously\cite{hb,hb1}.
 In the first case, the red shift of the hydrogen bond in N-H...O is from N-H bond length hardening with an associated decrease in the N-H stretching frequency and accompanied by an increase of the
 IR intensity. The same can be seen  in N22-H11 bond length and the frequencies in the mid-IR region (above 3000 cm$^{-1}$), which corresponds to NH$_2$ stretching motion.
 In the second case, the blue shifted hydrogen bond results from a shortening of the N-H bond lengths, and an increase in stretching frequency of NH group with decrease in IR intensity. A similar situation is
 noticed in Fig. 3, where IR spectra intensities from 530-1600 cm$^{-1}$ decreases and the corresponding mode frequencies are hardened with pressure. Also, N21-H12, N22-H21 and N22-H22
 bond lengths also decreases with pressure.  From these features, one can expect an increase of the hydrogen bond strength in FOX-7 under studied pressure range.

\begin{figure*}
\centering
\subfigure[]{\includegraphics[width=3.2in,clip]{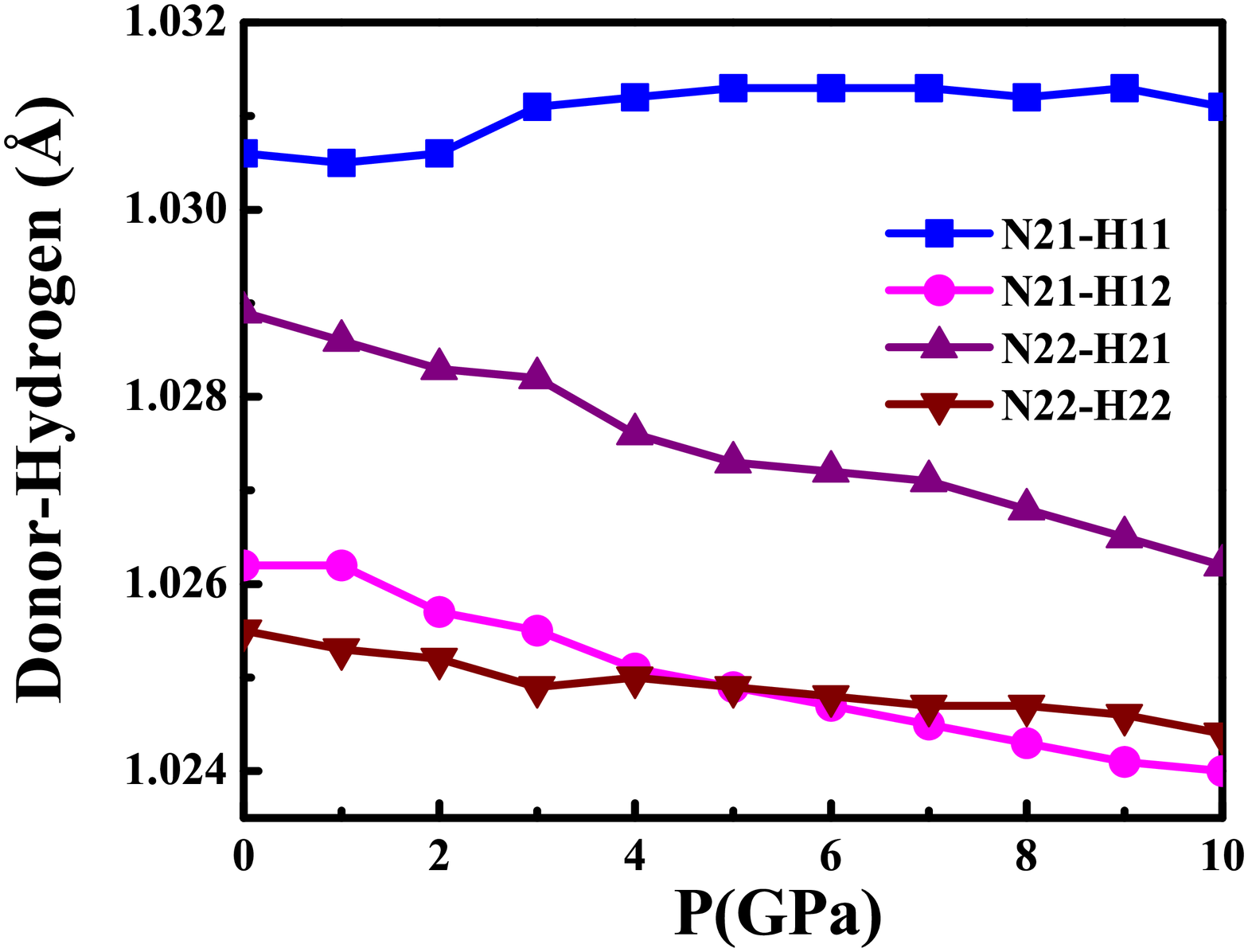}}
\subfigure[]{\includegraphics[width=3.2in,clip]{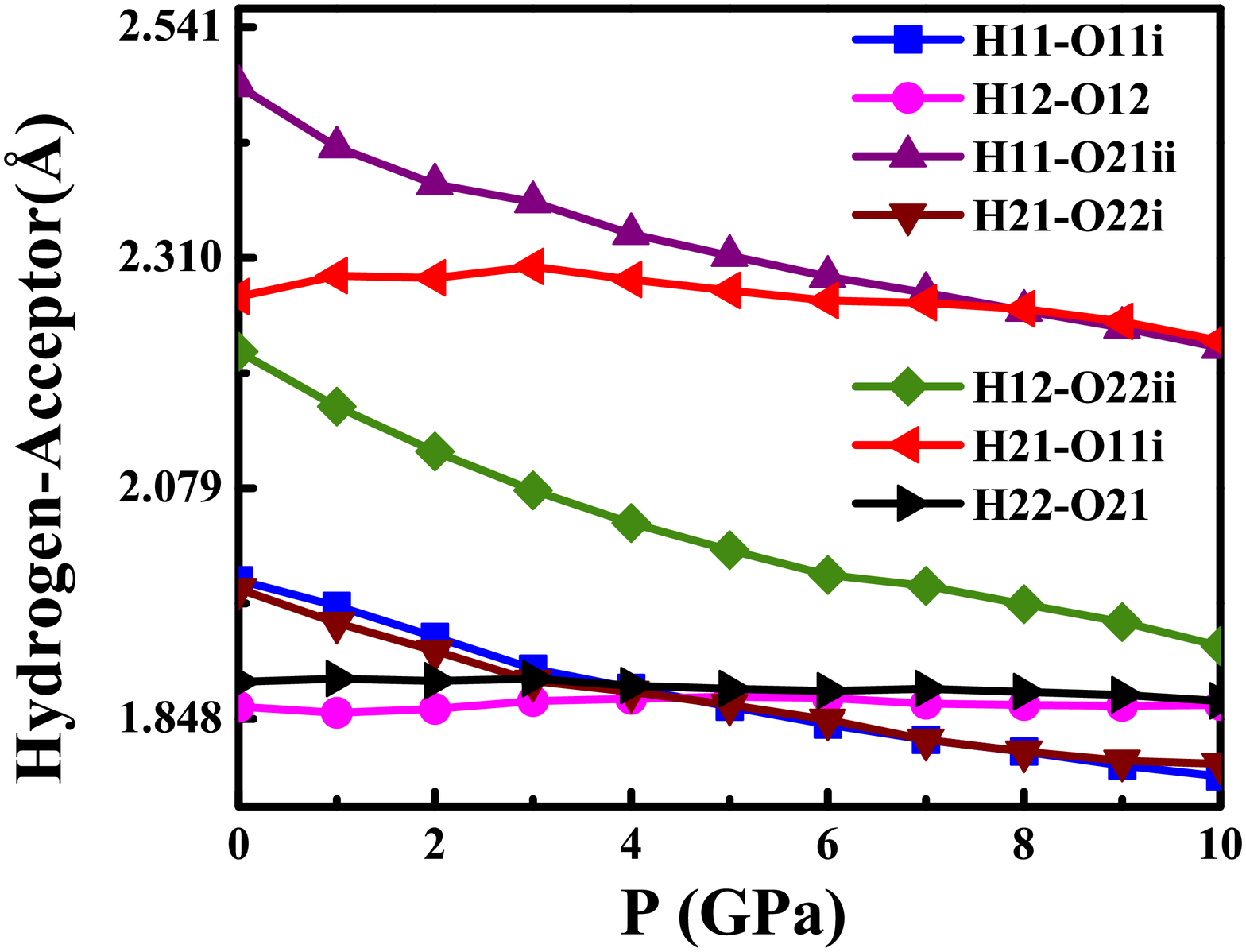}}
\subfigure[]{\includegraphics[width=3.2in,clip]{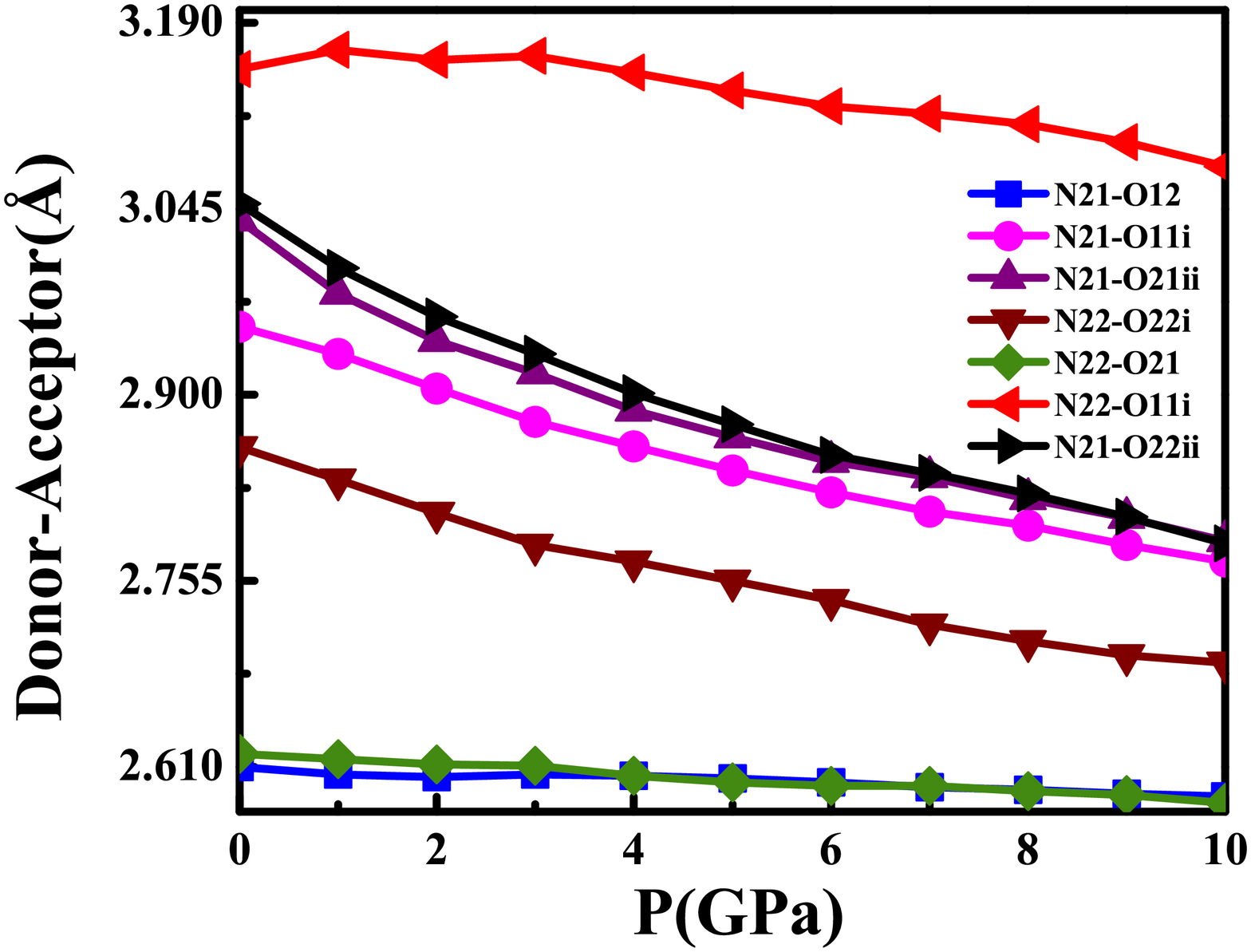}}
\caption{(Colour online) Calculated pressure dependence of inter- and intramolecular hydrogen bond lengths of FOX-7, (a). Donor-Hydrogen, (b). Acceptor-Hydrogen and (c). Acceptor-Donor. Note the different set of bond lengths were taken from experimental notation\cite{Ulf}. }
\end{figure*}


\subsection{Transition from $\alpha$ to t $\alpha'$ structure}

 As mentioned in the introduction, experiments indicates a phase transition from the $\alpha$ to the $\alpha'$ phase at pressure 2 GPa\cite{pravica,Bishop},
 (the $\alpha'$ phase was considered to be identical to the $\beta$ phase of FOX-7 found previously at 373 K). In contrast to this, recent experiment using
 isothermal compression (at 298 K) found a phase transition at the same pressure but they claimed that the structure was different from the $\beta$ structure\cite{gupta}.
 With the aim to have a better understanding of this phase transition, we have calculated the enthalpy for the $\alpha$ and $\beta$ phases taking into
 account Grimme's correction for vdw interactions. However we didn't observe any sign of a possible transition, and the $\alpha$ phase stayed with the lowest
  enthalpy for the range of pressure that we investigated. Therefore, we anticipate that temperature plays a major role during the transition, and/or that
   the level of theory (dipole-dipole correction) that we are using to take into account dispersive interaction is not sufficient.

\subsection{Quasiparticle band structure}

\begin{figure*}
\centering
{\includegraphics[width=6in,clip]{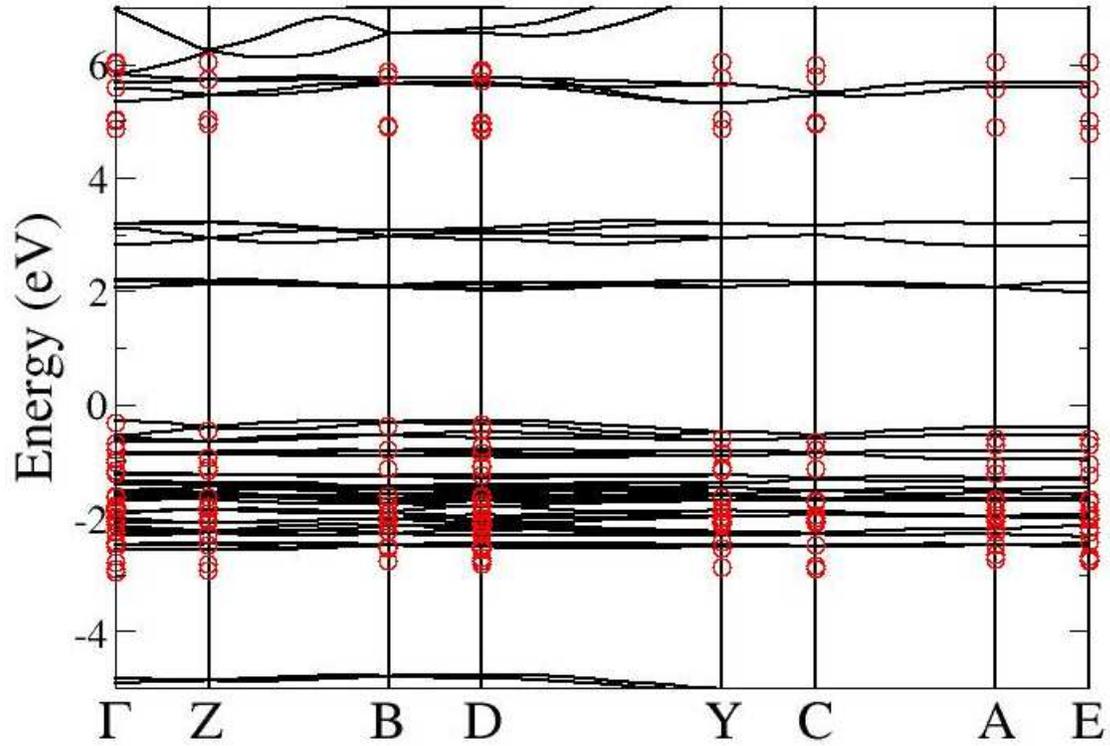}}
\caption{(Colour online)  GGA (black lines) and G$_0$W$_0$ (red circles) band structure of FOX-7 at experimental lattice parameters. Here , the high symmetry k points are $\Gamma$ : (0.0 0.0 00.0), Z: (0.0 0.5 0.0), B: (0.0 0.0 0.5), D: (0.0 0.5 0.5), Y: (0.5 0.0 0.0), C: (0.5 0.5 0.0), A: (-0.5 0.0 0.5) and E: (-0.5 0.5 0.5).}
\end{figure*}

Earlier, a few  theoretical studies using DFT have been made on metal azides  as well as
 organic explosives  to understand the correlation between the impact sensitivity and the band gap\cite{zhu},
   they claimed that a small band gap and  eases for electron transfer from the valence bands to the conduction bands, which leads eventually to decomposition and
  explosion. However, it is well known fact that the standard DFT band gaps are not adequate to predict band gaps.
  Nowadays,  several methods such as Hybrid functional like HSE\cite{Heyd,Krukau}, Tran and Blaha modified Becke Johnson potential\cite{Tran,david} and GW approximation\cite{Mel,Jones} are
  available to overcome this problem. These methods gives reasonable band gaps for various class of materials, but
  such studies were not addressed for the HEMs because of the crystal complexities and high computational time.
  Predicting a correct band gap is needed if related properties such as optical absorption, electronic band alignment for bonding,
  defect levels as well as sensitivity to light are studied.
  Since the experimental band gap of FOX-7 is not known, we choose to predict it with the help of the GW approximation\cite{seb1,seb2}
  in the  G$_0$W$_0$ flavor. The calculated band structures of FOX-7 with GGA
  and with the G$_0$W$_0$ approximation are shown in Fig. 6. From this, it is clearly observed that band profiles are similar for both methods, whereas the band gap increases with the G$_0$W$_0$ approximation. From the band structure, the top of the valence band and the bottom of the conduction
  band occur at the $\Gamma$ point with a band gap of 2.3 eV with GGA , which agree well  with earlier theoretical values\cite{Kuklja2,Sorescu}.
  The band gap obtained using the GW approximation is enhanced to 5.1 eV. A similar increase in the value of the band gap was obtained for solid nitromethane
  in some of our earlier work\cite{sa2}. Hence, it is essential to calculate exact band gap values for high energy materials by using advanced methods like the GW approximation or others.

\section{Conclusions}
 In conclusion, we have investigated the structural and vibrational properties in FOX-7 to understand the corresponding intra and intermolecular interactions under pressure.
 The ground state structural properties obtained with GGA+G06 shows an accurate agreement with experiments, and the calculations performed at high pressure
 shows a large compressibility along the b-axis, again in agreement with experiments.
 The vibrational properties computed under pressure show the a large contribution of intermolecular interactions takes place up to 10 GPa and demonstrate evidence of strong
 hydrogen bonding in the mid IR region. Finally, we studied the quasi particle band structure of FOX-7 to predict an accurate band gap and found an value of 5.1 eV,
 indicating the  wide band gap nature of FOX-7. We hope that our work will  stimulate further experimental studies, in particular concerning phase transitions in FOX-7 and
 its band gap.
\section{Acknowledgements}
\par S. A. would like to thank DRDO through ACRHEM for financial support. S. A. and G. V. thank CMSD, University of Hyderabad, for providing computational facilities.
S. L. acknowledge the acces to HPC resources from GENCI-CCRT/CINES (Grant x2014-085106).
$^*$\emph{Author for Correspondence, E-mail: gvaithee@gmail.com}


\begin {thebibliography}{}

\bibitem{Mathieu}
D. Mathieu, J. Phys. Chem. A {\bf116}, 1794 (2012).

\bibitem{Anniyappan}
M. Anniyappan, M. B. Talawar, G. M. Gore, S. Venugopalan, B. R. Gandhe  J. Hazard. Mater. {\bf137}, 812 (2006).


\bibitem{Nikolai}
N. V. Latypov, J. Bergman, A. Langlet, U. Wellmar, U. Bemm, Tetrahedron {\bf54}, 11525 (1998).


\bibitem{Ulf}
U. Bemm, H. \"{O}stmark, Acta. Cryst. C{\bf54}, 1997 (1998).



\bibitem{Peris}
S. M. Peiris, C. P. Wong, F. J. Zerilli, J. Chem. Phys  {\bf120}, 8060 (2004).

\bibitem{Kempa}
P. B. Kempa, M. Herrmann Part. Part. Syst. Charact. {\bf22}, 418 (2005).

\bibitem{Evers}
J. Evers, T. M. Klap\"{o}tke, P. Mayer, G. Oehlinger, J. Welhch Inorg. Chem {\bf45}, 4996 (2006).

\bibitem{Pravica}
M. Pravica, M. Galley, C. Park, H. Ruiz, J. Wojno, High Pressure Research  {\bf31}, 80 (2011).

\bibitem{Xue}
Xue-zhong Fan , Ji-zhen Li ,  Zi-ru Liu, J. Phys. Chem A, {\bf111}, 13291 (2007).

\bibitem{Gao}
Hong-Xu Gao, Feng-Qi Zhao, Rong-Zu Hu, Qin Pan, Bo-Zhou Wang, Xu-Wu Yang, Yin Gao, Sheng-Li Gao, Chin. Jou. Chem, {\bf24}, 177 (2006).

\bibitem{pravica}
M. Pravica, Y. Liu, J. Robinson, N. Velisavlijevic, Z. Liu, M. Galley, J. Appl. Phys., {\bf111}, 103534 (2012).

\bibitem{Bishop}
M. M. Bishop, R. S. Chellappa, M. Pravica,  J. Coe, Z. Liu, D. Dattlebaum, Y. Vohra N. Velisavlijevic,  J. Chem. Phys, {\bf137}, 174304 (2012).

\bibitem{gupta}
Z. A. Dreger, Y. Tao, Y. M. Gupta, Chem. Phys. Lett. {\bf584}, 83 (2013).

\bibitem{margaret}
Margaret-Jane Crawford, J\"{u}rgen Evers, Michael Göbel, Thomas M. Klap\"{ö}tke, Peter Mayer, Gilbert Oehlinger and Jan M. Welch, Propellants, Explosives, pyrotechnics {\bf32} 478 (2007).

\bibitem{Kuklja1}
M. M. Kuklja, E. V. Stefanovich, A. B. Kunz, J. Chem. Phys,  {\bf112}, 3417 (2003).

\bibitem{Kuklja2}
S. N. Rashkeev, M. M. Kuklja,  F. J. Zerilli, Appl. Phys. Lett,  {\bf82}, 1371 (2003).

\bibitem{Kuklja3}
 M. M. Kuklja,  F. J. Zerilli, S. M. Peiris  J. Chem. Phys,  {\bf118}, 11073 (2003).

\bibitem{Kuklja4}
 F. J. Zerilli,  M. M. Kuklja,  J. Phys. Chem A,  {\bf110}, 5173 (2006).

\bibitem{Kuklja5}
 M. M. Kuklja, S. N. Rashkeev,  F. J. Zerilli, Appl. Phys. Lett,  {\bf89}, 071904 (2006).

\bibitem{Kuklja6}
 M. M. Kuklja, S. N. Rashkeev,  Appl. Phys. Lett,  {\bf90}, 151913 (2007).

\bibitem{Kuklja7}
A. V. Kimmel, P. V. Sushko, A. L. Shluger, M. M. Kuklja,  J. Phys. Chem A,  {\bf112}, 4496 (2008).

\bibitem{Kuklja8}
 F. J. Zerilli,  M. M. Kuklja,  J. Phys. Chem A,  {\bf111}, 1721 (2007).

\bibitem{Kuklja9}
A. V. Kimmel, P. V. Sushko, A. L. Shluger, M. M. Kuklja,  J. Chem. Phys,  {\bf126}, 234711 (2007).

\bibitem{Kuklja10}
 M. M. Kuklja, S. N. Rashkeev, Phys. Rev. B,  {\bf75}, 104111 (2007).

\bibitem{Kuklja11}
 M. M. Kuklja, S. N. Rashkeev, J. Phys. Chem C Lett, {\bf113}, 17 (2009).

\bibitem{Kuklja12}
A. V. Kimmel, D. M. Ramo, P. V. Sushko, A. L. Shluger, M. M. Kuklja,  Phys. Rev. B,  {\bf80}, 134108 (2009).

\bibitem{Kuklja13}
 M. M. Kuklja, S. N. Rashkeev, Journal of Energetic Materials,  {\bf28}, 66 (2010).


\bibitem{Asta}
A. Gindulyt\.{e}, L. Massa, L. Huang, J. Karle, J. Phys. Chem. A {\bf103}, 11045 (1999).

\bibitem{Sorescu}
D. C. Sorescu, J. A. Boatz, D. L. Thompson J. Phys. Chem. A {\bf105}, 5010 (2001).

\bibitem{Hai}
Xue-Hai Ju, He-Ming Xiao and Qi-Ying Xia,  J. Chem. Phys,  {\bf119}, 10247 (2003).

\bibitem{Hu}
Anguang Hu, Brian Larade, Hakima Abou-Rachid, Louis-Simon Lussier and Hong Guo, Propellants, Explosives, Pyrotechnics  {\bf31}, 355 (2006).

\bibitem{Zhao}
J. Zhao, H. Liu, Computational Materials Science   {\bf42}, 698 (2008).

\bibitem {Zneng}
Z. Zheng, J. Xu, J. Zhao,  High pressure Research {\bf 30}. 301 (2010).

\bibitem {dftd}
D. C. Sorescu, B. M. Rice, J. Phys. Chem C {\bf 114}. 6734 (2010).

\bibitem {Landeville}
A. C. Landerville, M. W. Conroy, M. M. Budzevich, Y. Lin, C. T. White, I. I. Oleynik, Appl. Phys. Lett  {\bf 97}, 251908 (2010).

\bibitem{Payne}
M. C. Payne, M. P. Teter, D. C.  Allan, T. A. Arias and J. D. Joannopoulos, Rev. Mod. Phys. {\bf 64}, 1045 (1992).

\bibitem{Segall}
M. D. Segall, P. J. D. Lindan, M. J. Probert, C. J. Pickard, P. J. Hasnip, S. J. Clark and  M. C. Payne, J. Phys. Cond. Matt. {\bf 14}, 2717 (2002).

\bibitem{Vanderbilt}
D. Vanderbilt, Phys. Rev. B {\bf 41}, 7892 (1990).


\bibitem{PPerdew}
J. P. Perdew and A. Zunger, Phys. Rev. B {\bf 23}, 5048 (1981).

\bibitem{Perdew}
J. P. Perdew, K. Burke and M. Ernzerhof,  Phys. Rev. Lett. {\bf 77}, 3865 (1996).

\bibitem{PW91}
J. P. Perdew and Y. Wang, Phys. Rev. B {\bf45}, 13244 (1992).

\bibitem{Grimme}
S. Grimme, J. Comp. Chem. {\bf 27}, 1787 (2006).

\bibitem{TS}
A. Tkatchenko and M. Scheffler, Phys. Rev. Lett. {\bf102}, 073005 (2009).

\bibitem{OBS}
F. Ortmann, F. Bechstedt, and W. G. Schmidt, Phys. Rev. B {\bf73}, 205101 (2006).

%
%

\bibitem{Monkhorst}
H. J. Monkhorst and J. Pack,  Phys. Rev. B {\bf 13}, 5188 (1976).

\bibitem{Hedin1}
L. Hedin, Phys. Rev. {\bf 139} A796 (1965).

\bibitem{Hedin2}
 L. Hedin and S. Lundquist, Solid State Physics, edited by H. Ehrenreich, F. Seitz, and D. Turnbull, vol 23, Academic, New York (1969).

\bibitem{seb1}
S. Leb\`egue, B. Arnaud, M. Alouani, P. E. Bl\"ochl, Phys. Rev. B {\bf 67}, 155208 (2003).

\bibitem{seb2}
R. H. Scheicher, D. Y. Kim, S. Leb\`egue, B. Arnaud, M. Alouani, R. Ahuja,
Appl. Phys. Letters, {\bf 92},  201903 (2008).

\bibitem{Kresse}
G. Kresse, J. Furthm\"{u}ller, Phys. Rev. B  {\bf 54}, 11169 (1996).


\bibitem{sa2}
S. Appalakondaiah, G. Vaitheeswaran, S. Leb\`egue, J. Chem. Phys. {\bf138}, 184705 (2013).

\bibitem{hb}
J. Joseph, Eluvathingal D. Jemmis, J. Am. Chem. Soc., {\bf129}, 4620  (2007).

\bibitem{hb1}
X. Li, Lei Liu,  H. B. Schlegel, J. Am. Chem. Soc., {\bf124}, 9639 (2002),

\bibitem{zhu}
W. Zhu and H. Ziao, Struct. Chem. {\bf21}, 657-655 (2010).

\bibitem{Heyd}
J. Heyd, G. Scuseria, and M. Ernzerhof, J. Chem. Phys. {\bf118}, 8207 (2003)

\bibitem{Krukau}
A. V. Krukau, O. A. Vydrov, A. F. Izmaylov, and G. Scuseria, J. Chem. Phys. {\bf125}, 224106 (2006)

\bibitem{Tran}
F. Tran,  P. Blaha,  Phys. Rev. Lett, {\bf102}, 226401, (2009).

\bibitem{david}
D. Koller,  F. Tran,  P. Blaha,  Phys. Rev. B  {\bf83}, 195134, (2011).

 \bibitem{Mel}
John P. Perdew, Mel Levy, Phys. Rev. Lett {\bf 51}, 1884-1887 (1983).

\bibitem{Jones}
 R. O. Jones,  O. Gunnarsson, Rev. Mod. Phys  {\bf 61}, 689-746 (1989).


















\end {thebibliography}
 \end{document}